\documentclass[aps,pra,twocolumn,showpacs,superscriptaddress,groupedaddress]{revtex4}
\usepackage{graphicx} % Required for inserting images
\usepackage{amsmath}

\begin{document}

\title{Collective dynamics of harmonically trapped 1D quantum droplets under linear gravitational-like confinement}

\author{Saurab Das}
\affiliation{Indian Institute of Information Technology Vadodara, Gujarat, India 382 028}
\author{Jayanta Bera}
\affiliation{C. V. Raman Global University, Bhubaneswar, Odisha 752 054, India}
\author{Ajay Nath}
\affiliation{Indian Institute of Information Technology Vadodara, Gujarat, India 382 028}

%\keywords{Keyword1, Keyword2, Keyword3}

\begin{abstract} 
We investigate the dynamics of harmonically confined quantum droplets in a binary Bose–Einstein condensate within the one-dimensional extended Gross–Pitaevskii framework, including Lee–Huang–Yang corrections, under a constant linear (gravitational-like) potential. By analyzing the center-of-mass (COM) and width dynamics, we show that the monopole (breathing) mode remains governed by the harmonic confinement, with a frequency asymptotically insensitive to the linear perturbation, demonstrating the robustness of internal collective excitations against uniform external forcing. In contrast, the COM exhibits interaction-dependent transport, with weak confinement producing large susceptibility and rapid displacement, whereas strong confinement suppresses transport even under large forcing. The COM response decreases monotonically with increasing trap frequency. We further characterize the evolving quantum state through the quantum Fisher information and Wigner quasi-probability distributions, showing that the linear potential enables controlled generation of states with enhanced metrological sensitivity over finite times, while stronger confinement shifts the onset of the high-sensitivity regime to larger forcing strengths. Numerical simulations based on the split-step Fourier method confirm the dynamical stability of the obtained solutions. These results illustrate the impact of linear gravitational like trap on the collective excitations, transport, and quantum metrological properties of 1D ultradilute quantum fluids.
\end{abstract}
\maketitle

\section{Introduction} 
The emergence of ultradilute liquid-like quantum droplets (QDs) in weakly interacting Bose–Bose mixtures stands as one of the most striking manifestations of beyond-mean-field (BMF) physics \cite{Malomed1,Luo}. In Bose–Einstein condensates (BEC), QDs originate from a subtle competition between attractive effective mean-field (EMF) interactions and repulsive quantum fluctuations, the latter captured by the Lee–Huang–Yang (LHY) correction \cite{Petrov,Petrov1}. While in three-dimensional (3D) systems the LHY contribution counteracts mean-field collapse by providing a repulsive stabilization, in reduced one-dimensional (1D) geometries it instead manifests itself as an effective attraction, enabling the formation of self-bound states even without external trapping \cite{Parisi,Giorgini}. Since Petrov’s seminal theoretical prediction, QDs have been experimentally realized on a wide range of platforms, including hyperfine states binary BEC mixtures \cite{Cabrera,Semeghini}, heteronuclear Bose–Bose mixtures \cite{Cavicchioli,Errico,Skov}, and dipolar Bose gases stabilized by quantum fluctuations \cite{Schmitt,Barbut,Wenzel,Chomaz,Tanzi}. These systems exhibit a rich phenomenology that includes soliton–droplet crossovers \cite{Cappellaro,Cheiney}, vortex droplet formation and stability \cite{Kartashov,Li,Kartashov1,Lee,Otajonov1}, liquid–gas transitions \cite{He}, dimensional crossovers \cite{Zin}, collective excitations \cite{Tylutki,Liu,Astrakharchik,Sturmer,Dong,Otajonov2}, and thermal effects on droplet stability \cite{Guebli}. 

External confinement plays a crucial role in stabilizing and probing the properties of QDs. In ultracold-atom experiments, harmonic trapping remains the most widely used method for creating, stabilizing, and manipulating droplets \cite{Cabrera,Semeghini,Cheiney}, with theoretical studies showing its strong influence on stability thresholds, phase diagrams, and excitation spectra \cite{Guo}. Variational analyses of 1D Bose–Bose mixtures under harmonic confinement predict distinct differences in collective-mode frequencies between soliton-like and droplet-like regimes \cite{Cappellaro}, while 3D simulations in isotropic traps demonstrate significant modifications to droplet–gas transitions and excitation modes \cite{Liu}. Beyond mean-field LHY corrections, intercomponent correlations under confinement further reshape density distributions and dynamical relaxation in heteronuclear mixtures \cite{Kevrekidis}. The role of harmonic traps has also been systematically examined in contexts such as particle-number imbalance \cite{Flynn}, soliton–droplet crossovers \cite{Pathak,Li}, breathing-mode oscillations \cite{Liu,Zezyulin}, transport under time-dependent modulation \cite{Das}, and robustness against perturbations \cite{Bhatia}, collectively highlighting the decisive role of trap frequency and driving protocols in governing droplet geometry, coherence, and center-of-mass dynamics \cite{Katsimiga,Katsi}. In particular, harmonic confinement has been shown to shift stability boundaries, enable controlled fragmentation and merging, and serve as a powerful tool for transport and precision-sensing applications. However, despite these advances, considerably less attention has been given to the combined influence of harmonic confinement and linear gravitational-like potentials on QDs. While harmonic traps can tune droplet size, coherence, and collective excitations, and linear potentials can induce drift and trajectory deviations, their interplay remains largely unexplored. This unexplored regime is of fundamental interest, as it may reveal novel stability boundaries, transport mechanisms, and collective excitations with direct implications for precision gravimetry, quantum interferometry, and analog gravity simulations using ultracold droplets.

Here, we investigate the dynamics of harmonically confined quantum droplets in a binary BEC within the 1D extended Gross–Pitaevskii equation (eGPE)framework, including LHY corrections, in the presence of a constant linear gravitational-like potential. The reduced dimensionality provides access to strongly correlated and beyond-mean-field regimes, where the interplay between harmonic and linear confinements significantly influences droplet stability, coherence, and emergent ordering. To analyze the coupled dynamics, we employ an ansatz-based analytical treatment of the 1D eGPE and estimate the form of wavefunction along with BMF/EMF nonlinearities. In the purely harmonic case, the system obeys Kohn invariance \cite{Dobson,Kohn} with COM motion and a breathing mode set solely by the trap frequency.  We show that upon introducing a linear potential, this invariance is broken: while the monopole (breathing) mode remains essentially locked to the harmonic confinement and largely insensitive to the linear perturbation, the COM dynamics acquire an interaction-dependent transport character governed by a trap-frequency-controlled susceptibility. We identify distinct mobility regimes: weak confinement yields large COM susceptibility and rapid displacement toward the trap boundary even under small forcing, whereas strong confinement suppresses transport despite large applied forces. Accordingly, the rate of COM shift decreases monotonically with increasing trap frequency. We further quantify the phase-space structure of the evolving state via quantum Fisher information and Wigner quasi-probability distributions, demonstrating that linear forcing enables controlled generation of states with enhanced metrological sensitivity over finite times. This provides a theoretical framework for generating and preserving metrological resources, enabling time-robust quantum metrology schemes in 1D harmonically trapped QDs. Supported by split-step Fourier simulations, these results provide a unified and robust framework for understanding symmetry breaking, collective dynamics, and transport on harmonically trapped QDs under constant linear gravitational like confinement. 

This work is organized as follows. In Section II, we introduce the droplet setting, detailing the external harmonic confinement combined with static gravitational-like potentials, and establish the reduced single-component eGPE framework. Section III presents the analytical characterization of the COM dynamics of the droplet, demonstrating its consistency with Kohn’s theorem and unveiling the modulation effects induced by the linear potential. In section IV we explore the dynamical coherence of the droplet using quantum information–theoretic measures, quantum Fisher information, and the Wigner quasi-probability distribution, thereby providing phase-space diagnostics of static linear driving. Numerical validations of the analytical predictions are presented in Section V, underscoring their robustness and stability. Finally, Section VI concludes with a summary of the main results and an outlook on future research directions.

\section{Droplet Framework and Trapping Environment} 
We consider an elongated 1D homonuclear bosonic mixture with equal masses $m_{1} = m_{2} \equiv m$ trapped in external trap of the form,
\begin{equation}
V(x)=\frac{1}{2}M^{2}x^{2}-ax,
\label{eq:QD_1}
\end{equation}

with time-dependent intracomponent repulsion 
($g_{\uparrow\uparrow} (t)= g_{\downarrow\downarrow}(t) \equiv g(t) > 0$), and inter-component attraction ($g_{\uparrow\downarrow}(t) < 0$), a setting known to support self-bound droplet states under the condition $\delta g(t) = g_{\uparrow\downarrow}(t) + g(t) > 0$~\cite{Petrov, Petrov1}. The interaction strengths are experimentally tunable through control of the three-dimensional scattering lengths, achievable either by means of magnetic Fano-Feshbach resonances or via confinement-induced resonances, which can be accessed by adjusting the transverse confinement. The mixture consists of two equally populated hyperfine states ($N_{1} = N_{2} \equiv N$), rendering them dynamically equivalent. A promising candidate system is provided by the hyperfine states $\lvert F = 1, m_{F} = -1 \rangle$ and $\lvert F = 1, m_{F} = 0 \rangle$ of $^{39}\mathrm{K}$, which have already been employed in relevant three-dimensional experiments~\cite{Cabrera, Cheiney, Semeghini}. The system is confined in a harmonic trap with frequency $\omega_{x} \ll \omega_{\perp}$, ensuring its effective one-dimensionality, since strong transverse confinement along the $y$ and $z$ directions kinematically freezes motion outside the longitudinal $x$-axis with $M$ and $a$ representing harmonic oscillator frequency and strength of linear gravitational like trap, respectively. Within this setting, the dynamics of the droplet wave function $\Psi(x,t)$ is governed by the 1D eGPE~\cite{Petrov, Astrakharchik}:

\begin{equation}
i\hbar \frac{\partial \Psi}{\partial t} 
= -\frac{\hbar^{2}}{2m} \frac{\partial^{2}\Psi}{\partial x^{2}} 
+ \frac{\delta g(t)}{2} \lvert \Psi \rvert^{2} \Psi
- \frac{\sqrt{m}}{\pi \hbar} g^{3/2}(t) \lvert \Psi \rvert \Psi 
+ V(x)\Psi,
\label{eq:eGPE_dim}
\end{equation}

where the second term accounts for repulsive cubic mean-field interactions, 
while the third term encodes attractive BMF quantum LHY fluctuations, 
scaling as $\lvert \Psi \rvert^{2}$, which crucially stabilize droplet formation~\cite{Petrov, Astrakharchik}.  

To facilitate analysis, we rescale time, length, and the wave function in units of 
$\omega_{\perp}^{-1}$, $\sqrt{\hbar/(m\omega_{\perp})}$, and $(m\omega_{\perp}/\hbar)^{1/4}$, respectively. 
Likewise, the height $V_{0}$ and width $\sigma$ of the external potential are expressed in units of 
$\hbar \omega_{\perp}$ and $\sqrt{\hbar/(m\omega_{\perp})}$. 
This procedure yields the dimensionless eGPE,

\begin{equation}
i \frac{\partial \psi}{\partial t} 
= -\frac{1}{2}\frac{\partial^{2}\psi}{\partial x^{2}} 
+ g_{2}(t) \lvert \psi \rvert^{2} \psi
- g_{1}(t) \lvert \psi \rvert \psi
+ V(x)\psi,
\label{eq:QD_2}
\end{equation}

where the effective nonlinear coefficients are given by  

\begin{equation}
g_{2}(t) = \frac{\delta g(t)}{2} \left( \frac{m}{\hbar^{3}\omega_{\perp}} \right), 
\qquad 
g_{1}(t) = \frac{1}{\pi} \left( \frac{m g^{2}(t)}{\hbar^{3}\omega_{\perp}} \right)^{3/4}.
\label{eq:coeffs}
\end{equation}

Here, a normalization condition is imposed on the wavefunction $\int_{-\infty}^{+\infty} |\Psi(x,t)|^2dx=N$, where $N$ is the total number of condensate atoms. $g_{2}(t)$ characterizes the effective mean-field contribution, while $g_{1}(t)$ quantifies the strength of the LHY quantum correction. The balance between these two nonlinearities governs the emergence, stability, and dynamics of 1D quantum droplets. 

We consider the following wavefunction solution form of equation (\ref{eq:QD_2}) for the external trap chosen in equation (\ref{eq:QD_1}): 
\begin{equation}
 \Psi(x,t)= \sqrt{\sec(\sqrt{2}Mt)} F[X(x,t)]e^{i \theta(x,t)},  
\end{equation}\label{eq:QD_3}
where
\begin{equation}
 F[X(x,t)]=\frac{3 \mu}{G_{1}}\frac{ 1  }{1+\sqrt{1-\frac{\mu}{\mu_{0}} \frac{ G_{2}}{ G_{1}}  } \cosh[\sqrt{-\mu} X(x,t)]},    
\end{equation}\label{eq:QD_4}

\begin{eqnarray}
 \theta(x,t)=-\sqrt{2}M \tan(\sqrt{2}Mt) (1+x) x \nonumber\\
 -\sec^2(\sqrt{2}Mt)\left[ \frac{\mu }{2} + \frac{a^2  \tan^2(\sqrt{2}Mt)}{2M^2}\right] t,    
\end{eqnarray}\label{eq:QD_5}
and 
\begin{eqnarray}
 X(x,t)=\sec(\sqrt{2}Mt)x-\frac{a \sec(\sqrt{2}Mt)}{2M^2},    
\end{eqnarray}\label{eq:QD_5a}

with $\mu_{0}=-2/9$, $ \mu<0$, $G_1<0$, and $G_2>0$. Here, $-\frac{a }{2M^2}$ represents the COM.  On substitution of equation (\ref{eq:QD_3}) in the equation (\ref{eq:QD_2}), this results into:
\begin{equation}\label{eq:QD6}
 g_{2}(t) =\frac{G_{2} \sec(\sqrt{2}Mt)}{2} , \;\;\;\;g_{1}(t) =\frac{G_{1} \sec(\sqrt{2}Mt)^\frac{3}{2}}{2},
\end{equation}
\begin{equation}\label{eq:QD7}
-\frac{\partial^2 F}{\partial X^2}-G_1\mid F(X)\mid F(X) +G_2\mid F(X)\mid^{2} F(X)=\mu F(X). 
\end{equation}
Here, $G_{1}$ and $G_{2}$ are strengths of BMF and EMF interactions, whereas $\mu$ represents the chemical potential of equation (\ref{eq:QD7}). It is apparent from the above that the phase ($\theta(x,t)$), BMF $(g_{1}(x,t))$, and EMF $(g_{2}(x,t))$ depend on the harmonic oscillator frequency ($M$). 

Here, we would like to highlight two previously investigated scenarios:\\
(i) free space: for $M=0$, and $a=0$, from equation (\ref{eq:QD_1}), the resultant external trap becomes $V(x,t)=0$ i.e. free space, and correspondingly the wavefunction takes the form $\Psi(x,t)=F[X]exp(-i \mu t)$ reported by Petrov \cite{Petrov}. \\
(ii) harmonic trap: for $a=0$, the external trap and the wavefunction are reduced to the case of the harmonic trap studied earlier \cite{Tylutki,Pathak}.

\begin{figure}
\centering
\includegraphics[scale=0.5]{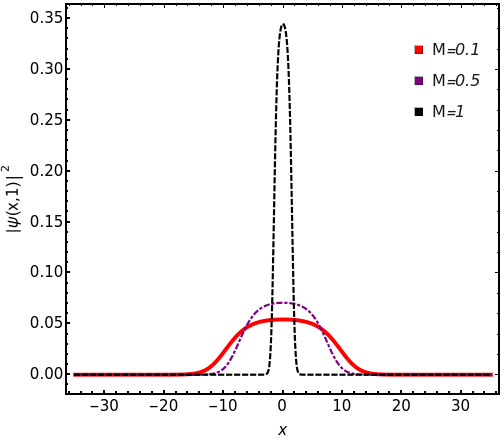}
\caption{\label{fig1} Density profiles for (a) harmonic trap: $t=1$, $\mu=-2/9$, $G_{1} = -1$, $G_{2} = 0.9999$, along with $M = 0.1$ (red thin line), $0.5$ (purpole dotdashed), $1$ (black dashed); (b) harmonic trap along linear gravitational like trap: regular harmonic trap with $V(x,t)=\frac{1}{2}Mx^{2}$ (blue dotdashed line), and expulsive harmonic trap with $V(x,t)=-\frac{1}{2}Mx^{2}$ (red dashed line) for the physical parameter values: $\mu=-2/9$, $G_{1} = -1$, $M=0.4$, $t=1$, $G_{2}=0.9999$ and $\gamma_{0}=1.5$. It is clear that the presence of external trap leads to compression in droplet profile.}
\end{figure}

Using the wavefunction solution obtained in equation (\ref{eq:QD_5}), we investigate the condensate dynamics under the changing strength of linear gravitational like trap ($a$) in combination with harmonic confinement. Here, we bound the droplet in the domain $[-50,50]$ and total number of atoms ($N$) is scaled to $1$. We illustrate the variation in the COM and width of the droplet with $a$, $M$ and $G_{2}$ for a given time. We further quantify the phase-space structure of the evolving state via quantum Fisher information and Wigner quasi-probability distributions, demonstrating that linear forcing by modulation of $M$ enables controlled generation of states with enhanced metrological sensitivity over finite times.

\begin{figure*}
\centering
\includegraphics[scale=0.5]{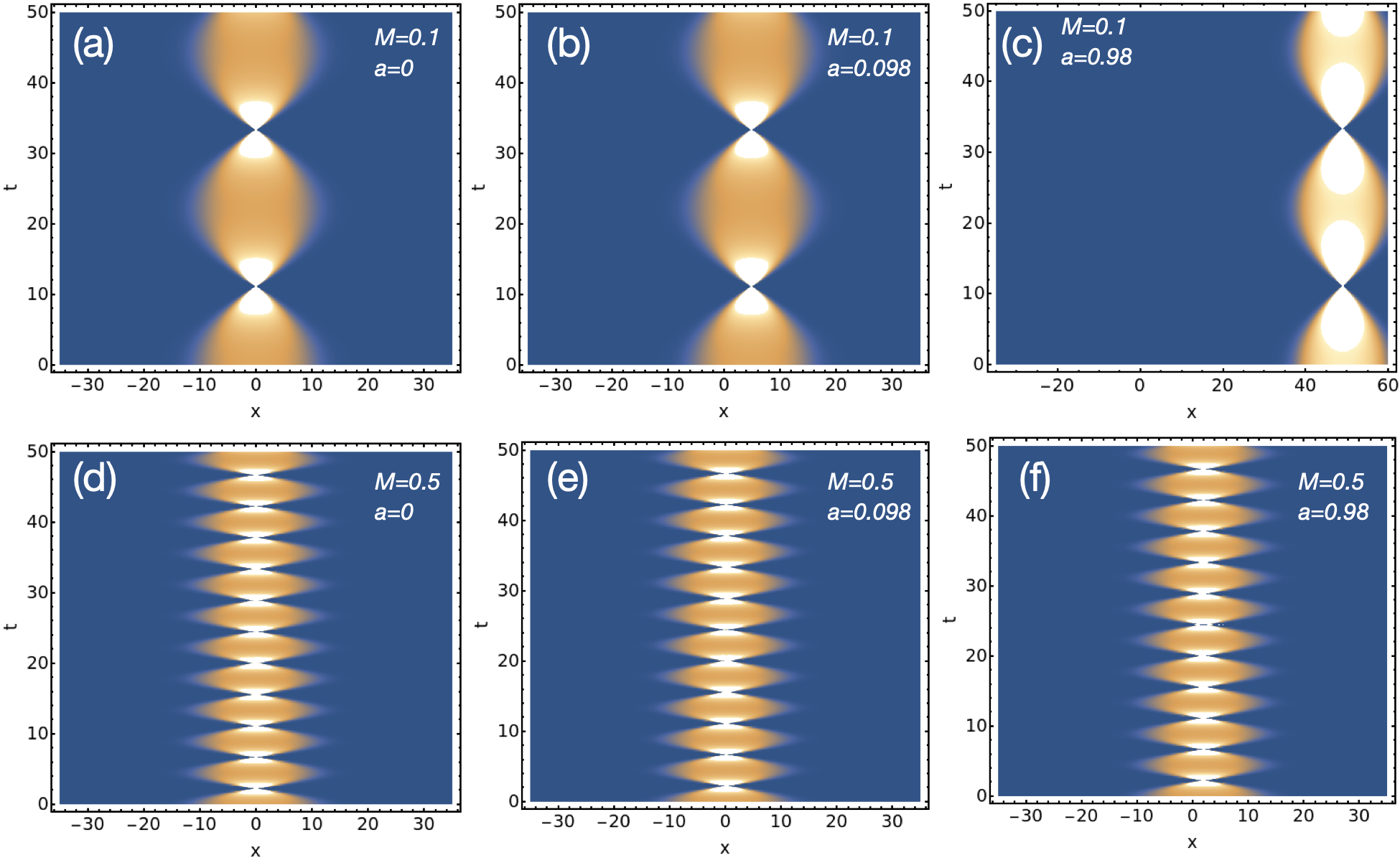}
\caption{\label{fig2} Temporal evolution of the QDs density with two different harmonic oscillator strength ($M$) under different gravitational-like acceleration strengths $a$. Subfigures (a)–(c) and (d)-(f) correspond to $a = 0$, $0.098$, $0.98$, for $M=0.1$, and $M=0.5$ respectively. The results illustrate the effect of $M$  and $a$ strengths on the dynamical response of the QD, including increase in breathing mode frequency. The simulations are carried out for the parameters $\mu = \mu_0 = -\frac{2}{9}$, $G_1 = -1$, and $G_2 = 0.9999$. The spatial coordinate is normalized by the harmonic oscillator length. .}
\end{figure*}

\section{Dynamics of harmonically trapped droplets under linear gravitational like trap }
We first consider the unperturbed harmonic confinement by setting the linear potential strength to $a=0$. The corresponding density distribution, $\rho=|\Psi|^2$, is shown in Fig.~\ref{fig1}, where the effect of the harmonic trap frequency $M$ is examined by varying it from $0.1$ to $1$ at $t=1$. As expected, increasing $M$ strengthens the harmonic confinement, resulting in a more localized quantum droplet. To further elucidate the spatiotemporal dynamics of the quantum droplet under the combined action of harmonic and linear potentials, the density evolution is presented in Fig.~\ref{fig2} for $M=0.1$ and $M=0.5$, with $a=0$, $0.098$, and $0.98$. For the weak-confinement case ($M=0.1$), the COM shifts from $x=0$ to approximately $x=5$ and $x=50$ as $a$ increases from $0$ to $0.098$ and $0.98$, respectively. In contrast, for $M=0.5$, the COM displacement remains comparatively small because the stronger harmonic confinement effectively suppresses the influence of the linear potential. These results demonstrate that the COM response is governed by the competition between the linear driving force and the harmonic restoring force, with the former becoming dominant only in the weak-confinement regime. In addition, the droplet width exhibits a periodic breathing motion whose frequency is determined solely by the harmonic confinement for a given value of $M$ and remains essentially unchanged with increasing $a$. This observation indicates that the monopole (breathing) mode is insensitive to the linear perturbation, consistent with the robustness of the internal compressional dynamics against a uniform external force. Increasing the trap frequency from $M=0.1$ to $M=0.5$ correspondingly increases the breathing-mode frequency.

%We begin with a unperturbed harmonic case by taking $a=0$ (i.e. no linear trap) and plot density ($\rho =|\Psi|^2$) in figure  (\ref{fig1}). First, we investigate the impact of harmonic oscillator frequency ($M$) strength by changing its magnitude from $0.1 \rightarrow 1$ in figure (\ref{fig1}). To further explore the spatio-temporal dynamics of QDs in linear and harmonic trap configuration, we do the density plot for $M=0.1$ and $M=0.5$ for $a=0, 0.098, 0.98$ in figure (\ref{fig2}). It is apparent from the figure that in comparison to $M=0.5$ case, the COM position of QDs shifts from $x=0$ location to $x=5$ and $x=50$ with $a$ changing from  $0 \rightarrow 0.098 \rightarrow 0.98$, respectively for the small magnitude of $M=0.1$. This is attributed to the fact that the strong confining harmonic trap strength suppresses the weak linear driving force whereas in the weak confining domain, the impact of even weak linear driving force significantly impact the COM location of QDs. Further, it is apparent from the figure that the width of droplet demonstrating constant breathing mode frequency for a particular $M$ magnitude and independent of change in $a$ value indicating that the monopole (breathing) mode remains fixed by the harmonic confinement, exhibiting a frequency that is asymptotically insensitive to the linear perturbation, reflecting the robustness of internal compressional dynamics against uniform external forcing. The increase in $M$ from $0.1 \rightarrow 0.5 $ leads to enhanced breathing mode frequency. 

\begin{figure*}
\centering
\includegraphics[scale=.43]{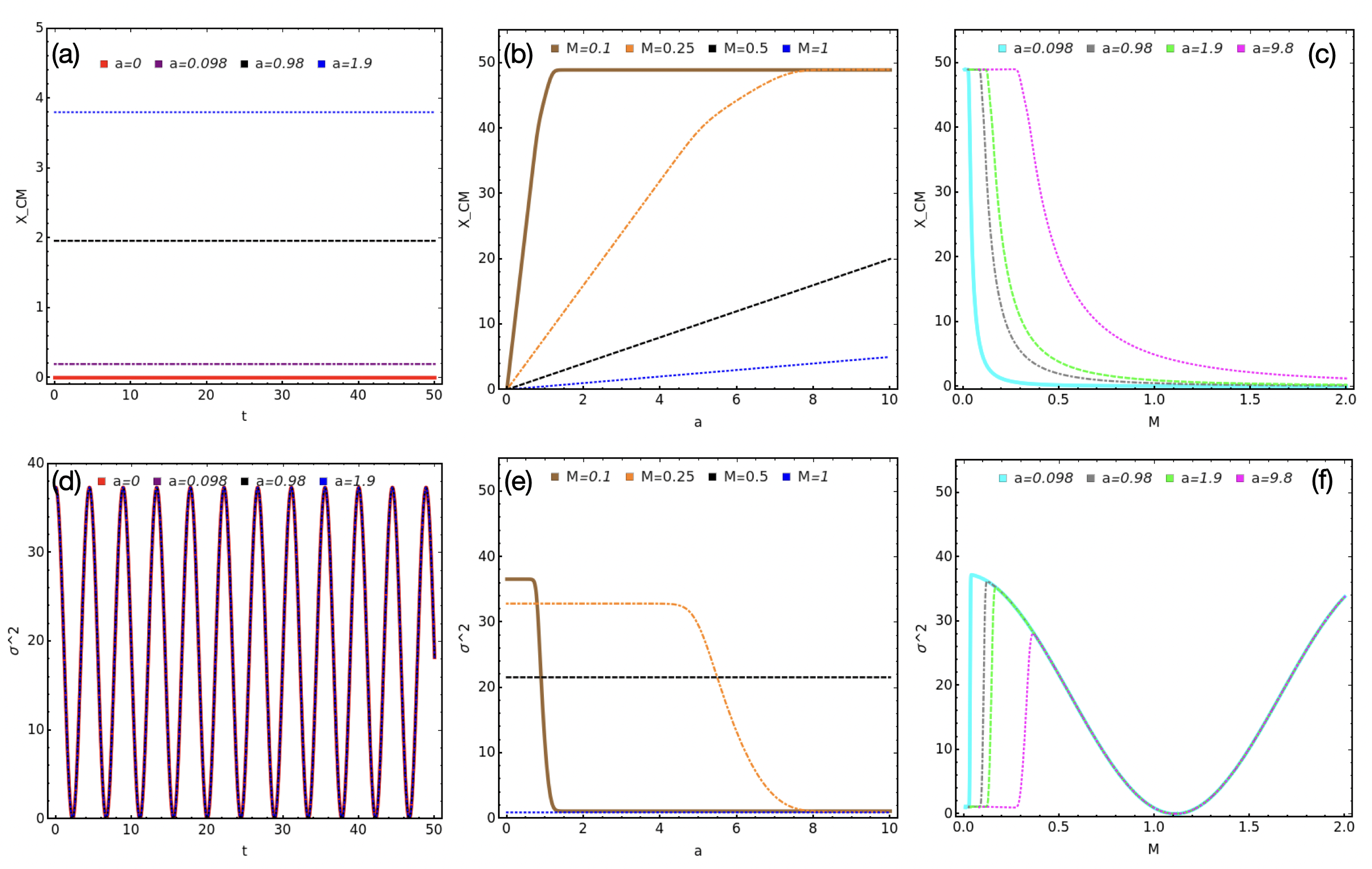}
\caption{\label{fig3}  Dependence of the COM position ($X_{\mathrm{CM}}$), and width ($\sigma^{2}$) of the QD on the linear gravitational-like potential strength $(a)$ and the harmonic trap frequency $M$. (a) Time evolution of $X_{\mathrm{CM}}$ and (d) $\sigma^{2}$ for $a=0$ (red solid), $0.098$ (purple dot-dashed), $0.98$ (black dot-dashed), and $1.9$ (blue dotted) with $M=0.5$. (b) $X_{\mathrm{CM}}$ and (e) $\sigma^{2}$ as functions of $a$ at $t=1$ for $M=0.1$ (brown solid), $0.25$ (orange dot-dashed), $0.5$ (black dashed), and $1$ (blue dotted). (c) $X_{\mathrm{CM}}$ and (f) $\sigma^{2}$ as functions of $M$ for $a=0.098$ (cyan solid), $0.98$ (gray dot-dashed), $1.9$ (green dashed), and $9.8$ (magenta dotted). The remaining parameters are fixed at $\mu=\mu_0=-2/9$, $G_1=-1$, and $G_2=0.9999$ .}
\end{figure*}

Next, we examine the influence of the linear gravitational-like potential strength ($a$) on the COM $(X_{CM})$ position and width ($\sigma^{2}$) of the QD, as shown in Fig.~\ref{fig3}. Unless otherwise stated, the parameters are fixed at $\mu=\mu_0=-2/9$, $G_1=-1$, and $G_2=0.9999$. Figures~\ref{fig3}(a) and \ref{fig3}(d) show the temporal evolution of the $X_{CM}$ position, and the $\sigma^2$, respectively, for $a=0$ (red solid), $0.098$ (purple dot-dashed), $0.98$ (black dot-dashed), and $1.9$ (blue dotted), with $M=0.5$. The COM is displaced to the equilibrium position $-a/(2M^2)$, demonstrating that its equilibrium location shifts linearly with the strength of the linear potential. In contrast, the width exhibits a breathing-mode oscillation with period $2\pi/M$, independent of $a$, indicating that the breathing dynamics are governed solely by the harmonic confinement, in agreement with the Kohn theorem.

Figures~\ref{fig3}(b) and \ref{fig3}(e) display $X_{\mathrm{CM}}$ and $\sigma^2$, respectively, as functions of $a$ at $t=1$ for $M=0.1$ (brown solid), $0.25$ (orange dot-dashed), $0.5$ (black dashed), and $1$ (blue dotted). For weak confinement ($M<0.5$), $X_{\mathrm{CM}}$ increases nonlinearly with $a$, whereas for $M\geq0.5$ the dependence becomes nearly linear, with the COM approaching the system boundary for sufficiently large $a$. The width decreases with increasing $M$, while its sensitivity to $a$ is pronounced only for $M<0.5$. In the strong-confinement regime ($M\geq0.5$), $\sigma^2$ is essentially unaffected by the linear potential.

Finally, Figs.~\ref{fig3}(c) and \ref{fig3}(f) present the dependence of $X_{\mathrm{CM}}$ and $\sigma^2$ on the trap frequency $M$ for $a=0.098$ (cyan solid), $0.98$ (gray dot-dashed), $1.9$ (green dashed), and $9.8$ (magenta dotted). As $M$ increases, the width rapidly converges to a common value for all $a$, confirming that the influence of the linear potential is restricted to the weak-confinement regime ($M<0.5$). Conversely, $X_{\mathrm{CM}}$ approaches the trap center with increasing $M$, although convergence becomes slower for larger $a$, reflecting the competition between the linear potential and the harmonic confinement. Thus, we identify distinct mobility regimes: weak confinement yields large COM susceptibility and rapid displacement toward the trap boundary even under small forcing, whereas strong confinement suppresses transport despite large applied forces.

\begin{figure*}
\centering
\includegraphics[scale=.45]{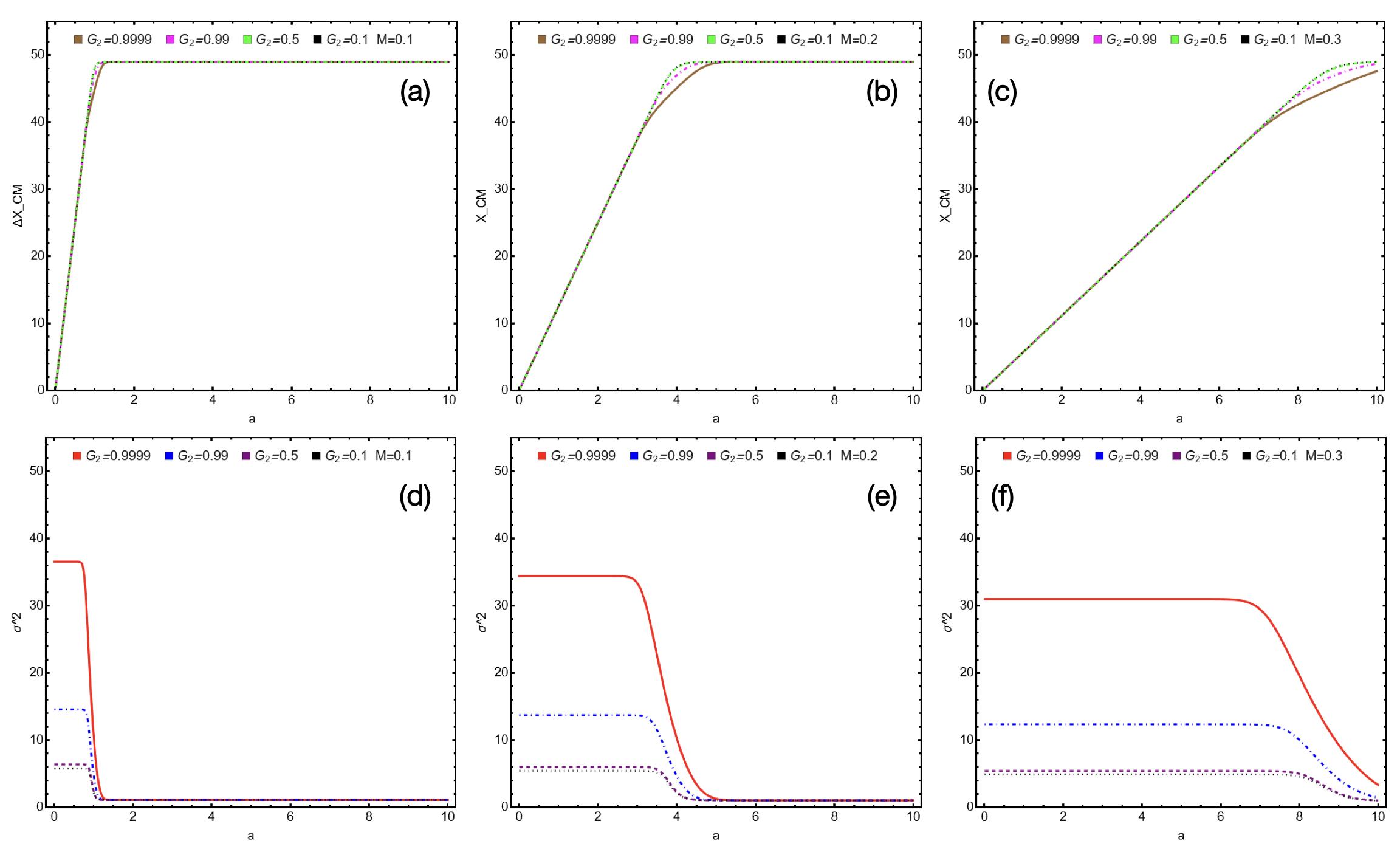}
\caption{\label{fig4} Dependence of the (a)–(c) center-of-mass position, $X_{\mathrm{CM}}$, and (d)–(f) droplet width, $\sigma^{2}$, on the EMF strength $G_{2}$ as functions of the linear gravitational-like trap strength $a$ for harmonic confinement strengths $M=0.1$ [(a),(d)], $M=0.2$ [(b),(e)], and $M=0.3$ [(c),(f)]. The curves correspond to $G_{2}=0.9999$ (brown, red solid), $0.99$ (magenta, blue dot-dashed), $0.5$ (green, purple dashed), and $0.1$ (black dotted) for both cases. Increasing $M$ shifts the EMF-sensitive regime to larger values of $a$, resulting in a stronger dependence of both $X_{\mathrm{CM}}$ and $\sigma^{2}$ on the EMF strength over a wider range of gravitational-like trap strengths. The remaining parameters are $t=1$, $\mu=\mu_{0}=-2/9$, and $G_{1}=-1$. .}
\end{figure*}

Additionally, we examine the influence of the EMF strength ($G_{2}$) on the $X_{\mathrm{CM}}$ and $\sigma^{2}$ as functions of the linear gravitational-like trap strength ($a$) for different harmonic confinement strengths ($M$), as shown in Fig.~\ref{fig4}. Unless otherwise stated, the parameters are fixed at $t=1$, $\mu=\mu_{0}=-2/9$, and $G_{1}=-1$. The reference value of the EMF strength is $G_{2}=0.9999$, except where it is varied explicitly. Figures~\ref{fig4}(a)–(c) depict $X_{\mathrm{CM}}$ for $M=0.1$, $0.2$, and $0.3$, respectively, with $G_{2}=0.9999$ (brown solid), $0.99$ (magenta dot-dashed), $0.5$ (green dashed), and $0.1$ (black dotted). As $M$ increases, the $X_{\mathrm{CM}}$ curves exhibit a progressively stronger separation, demonstrating an enhanced sensitivity to the EMF strength. The onset of this separation shifts toward larger trap strengths, occurring near $a\approx1$, $4$, and $8$ for $M=0.1$, $0.2$, and $0.3$, respectively.

The corresponding evolution of the droplet width is shown in Figs.~\ref{fig4}(d)–(f). For weak harmonic confinement ($M=0.1$), $\sigma^{2}$ is largest for $G_{2}=0.9999$ and decreases systematically with decreasing $G_{2}$ over the interval $0\le a\le2$, beyond which the curves nearly coincide. Increasing the harmonic confinement shifts this EMF-sensitive region to larger values of $a$, extending up to $a\approx4$ for $M=0.2$ and to approximately $a\approx10$ for $M=0.3$. Thus, stronger harmonic confinement not only delays the onset of the EMF-induced modifications in $X_{\mathrm{CM}}$ but also broadens the range of gravitational-like trap strengths over which the EMF significantly affects the droplet size.

%Additionally, we explore the impact of EMF strength ($G_{2}$) on the $X_{CM}$ and $\sigma^{2}$ with the increase in the linear gravitational like trap strength ($a$) at constant linear harmonic oscillator strength ($M$) in figure ~\ref{fig4}. In figure ~\ref{fig4} [(a)-(c)], the distinct $X_CM$ is observed for increasing magnitude of $M$ changing from $0.1 \rightarrow 0.3$ for $G_{2}=0.9999$ (brown thin line), $0.99$ (Magenta dotdashed), $0.5$ (green dashed), and $0.1$ (black dotted), respectively when we tune the strength of $a$. Further, this distinct $X_CM$ is dependent on the choice of $M$ and this distinct values are observed around $a=1$ ($M=0.1$), $a=4$ ($M=0.2$), and $a=8$ ($M=0.3$), respectively. Next, we depict the variation of $\sigma^2$ for $G_{2}=0.9999$ (red thin line), $0.99$ (blue dotdashed), $0.5$ (purple dashed), and $0.1$ (black dotted), respectively for $M$ changing from $0.1 \rightarrow 0.3$. For $M=0.1$, the droplet width is maximum for $G_{2} =0.9999$ which gradually decreases with decrease in $G_{2}$ in the range of $a$ changing from $0 \rightarrow 2$ and remain same for higher magnitude of $a$. However, as $M$ changes from $0.1 \rightarrow 0.2$, this variation shifts from $a = 0$ to $4$ and remain same for higher magnitude of $a$. Whereas for $M=0.3$, $\sigma^2$ is varying in between $a$ to $0-10$.

\begin{figure*}
\centering
\includegraphics[scale=.75]{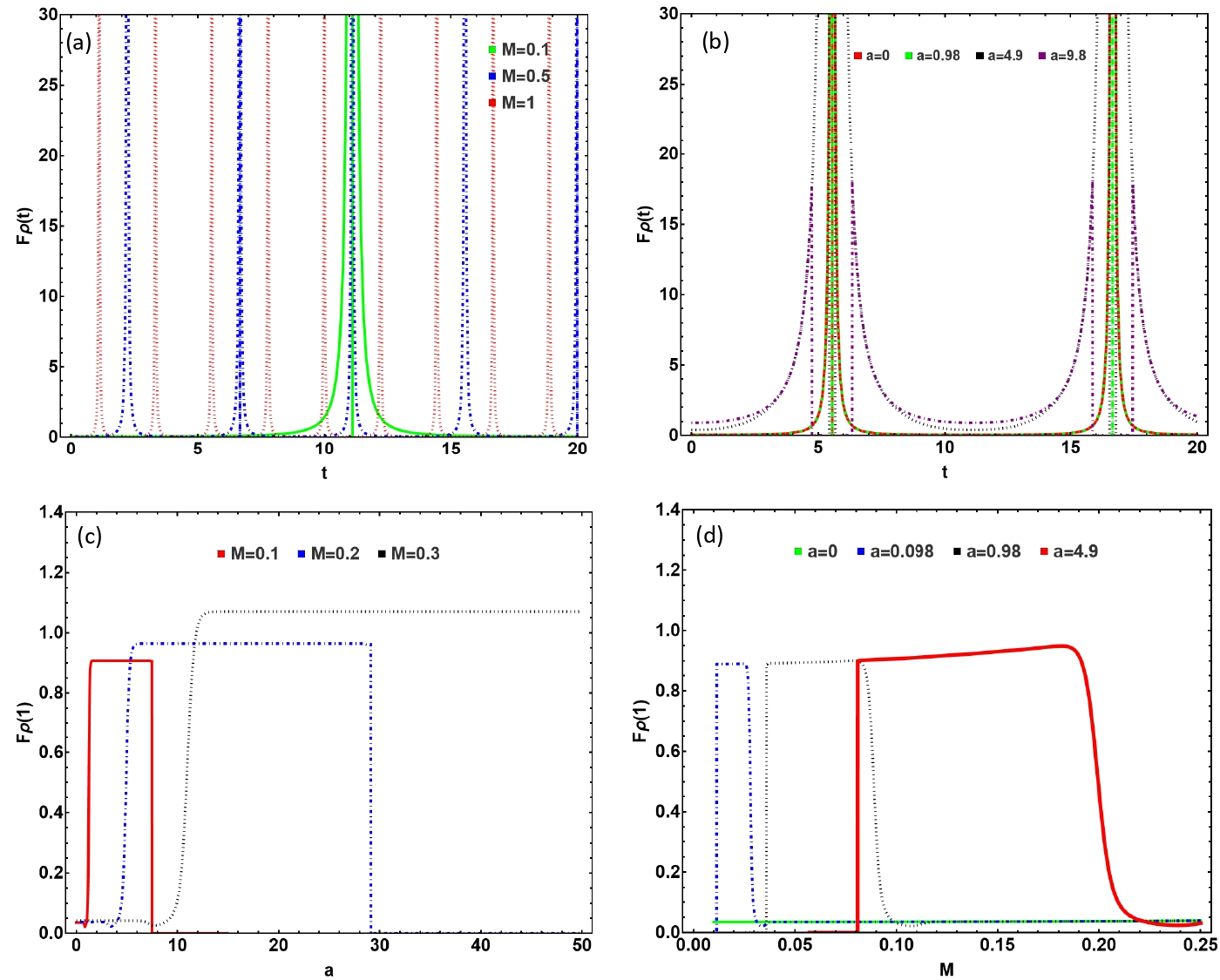}
\caption{\label{fig5} Quantum Fisher information $F_{\rho}(t)$ of the QDs. (a) Time evolution for a purely harmonic trap ($a=0$) with $M=0.1$, $0.5$, and $1$. (b) Time evolution for $M=0.2$ and different linear gravitational-like trap strengths $a$. (c) $F_{\rho}(1)$ as a function of $a$ for different $M$. (d) $F_{\rho}(1)$ as a function of $M$ for different $a$. The linear gravitational-like potential extends the temporal and parameter regimes of enhanced quantum sensitivity.}
\end{figure*}

\section{Quantum Fisher Information and Wigner Function}
\subsection{Quantum Fisher Information} The Quantum Fisher Information (QFI) defined as
\begin{equation}
F_{\rho}=\int \frac{1}{\rho(x,t)}
\left(\frac{\partial \rho(x,t)}{\partial x}\right)^2 dx,
\end{equation}

quantifies the sensitivity of a quantum state to variations of an unknown parameter and plays a central role in quantum metrology~\cite{Rath2021}. Unlike global information measures such as the Shannon or von Neumann entropy, the QFI is a local quantity that depends on spatial variations of the probability density. In many-body systems, it has been widely used to characterize multipartite entanglement, quantum correlations, and critical phenomena~\cite{Pezze2017}. More recently, information-theoretic measures, including the QFI, have been employed to distinguish structural transitions between flat-top and sharp-top quantum droplets~\cite{Siddik2024}. Here, under external linear driving, the collective modes of QDs are excited leading to dynamical modulation of QFI. 

Motivated from this, we investigate the QFI under weak harmonic for different strength of linear gravitational strength $a$. Figure~\ref{fig5}(a) shows the time evolution of $F_{\rho}(t)$ for a purely harmonic trap ($a=0$) with $M=0.1$, $0.5$, and $1$. The QFI exhibits periodic revivals characterized by sharp peaks separated by intervals of nearly vanishing values. These peaks occur when $\sqrt{2}Mt=(2k+1)\pi/2$ ($k \neq 0$) as wavefunction is dependent on $\sec(\sqrt{2}Mt)$, where the scaling factor diverges, corresponding to maximal localization of the droplet and, consequently, maximum parameter sensitivity that saturates the Cramér-Rao bound \cite{Escher}. Increasing the harmonic confinement reduces the revival period while preserving the overall dynamics. 

Figure~\ref{fig5}(b) illustrates the influence of $a$ for the fixed harmonic confinement ($M=0.2$). In the absence of the linear potential ($a=0$), the QFI displays narrow, isolated revivals. As the linear trap strength increases, these peaks broaden and the QFI remains finite over a larger time interval, indicating an extended temporal window of enhanced parameter sensitivity. 

To optimize parameter selection for the system, Figs.~\ref{fig5}(c) and ~\ref{fig5}(d) provide a complementary analysis of the QFI snapshot at a fixed time $t=1$, denoted as $F_\rho(1)$, revealing how the harmonic trapping frequency $M$ and the linear potential strength $a$ work in tandem. Figure~\ref{fig5}(c) shows that a minimum linear tilt $a$ is required to trigger the high-sensitivity plateau. Although tighter harmonic confinement shifts the onset of this plateau to higher threshold values, it simultaneously expands the stable operational range. In contrast, Fig~\ref{fig5}(d) highlights that a pure harmonic trap ($a=0$) keeps quantum sensitivity entirely suppressed. Introducing $a \neq 0$ leads to a well-defined window of enhanced precision over a specific span of $M$ which broadens with increase in magnitude of $a$. Thus, a finite linear potential is required to induce a high-QFI regime, while stronger harmonic confinement shifts its onset to larger $a$ and broadens the corresponding operational range. Likewise, increasing $a$ enlarges the range of harmonic confinement over which the QFI remains enhanced, demonstrating that the interplay between the harmonic and linear gravitational like confinements provides an effective means to control the metrological performance of the QDs.

\begin{figure*}
\centering
\includegraphics[scale=.32]{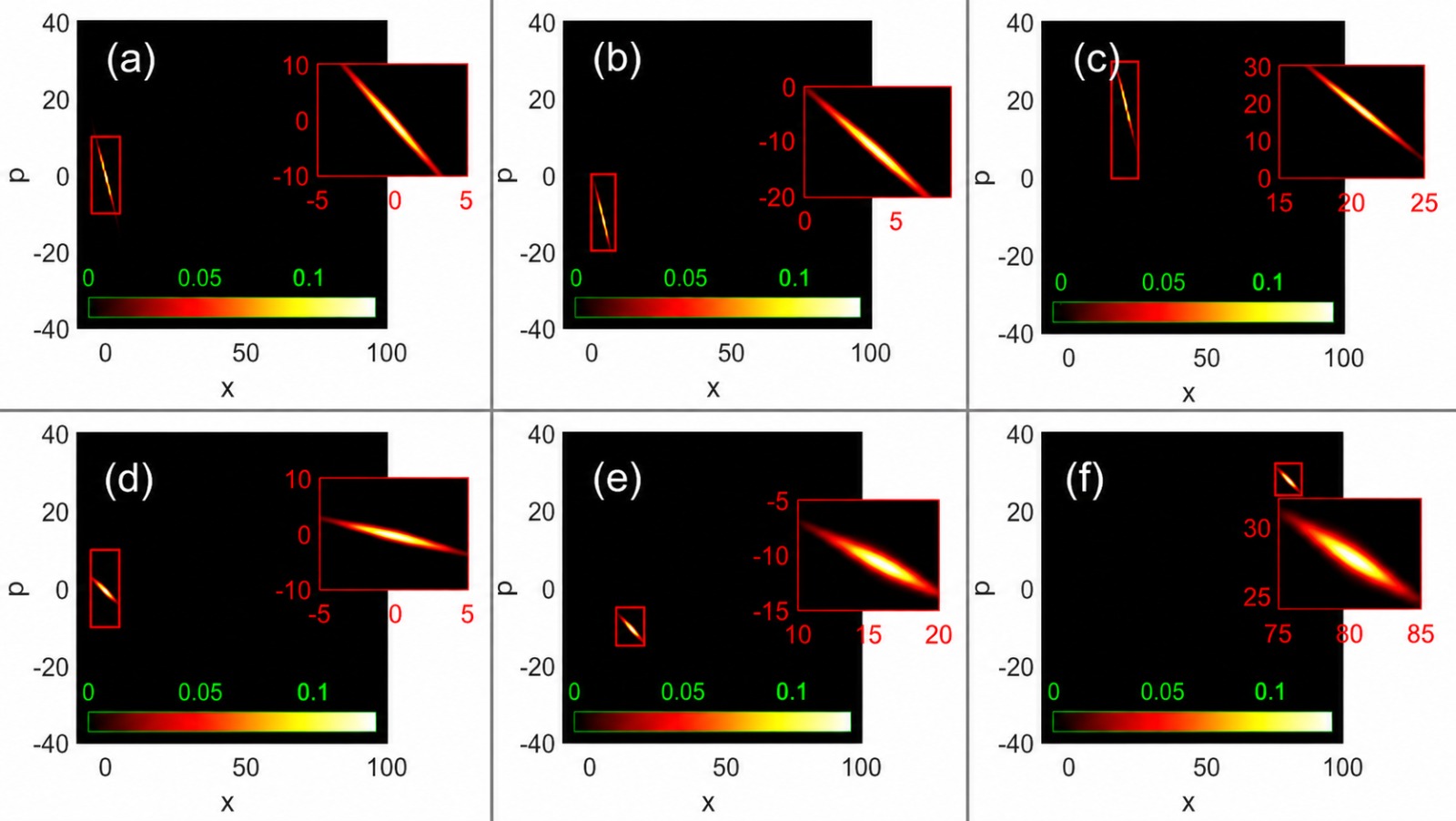}
\caption{\label{fig6} Wigner quasi-probability distribution $W(x,p)$ of the quantum droplet for linear gravitational-like trap strengths $a=0$ [(a),(d)], $2$ [(b),(e)], and $10$ [(c),(f)]. Panels (a)–(c) correspond to $M=0.5$, and panels (d)–(f) to $M=0.25$. Increasing $a$ displaces the phase-space distribution and modifies its momentum width, while weaker harmonic confinement enhances the spatial displacement and phase-space localization. Other parameters are $t=1$, $\mu=\mu_{0}=E=-2/9$, $G_{1}=-1$, and $G_{2}=0.9999$.}
\end{figure*}

\subsection{Wigners Phase-Space Distribution}

To characterize the phase-space dynamics of the QD and examine the effects of the linear gravitational-like potential and harmonic confinement, we evaluate the Wigner quasi-probability distribution, $W(x,p)$ \cite{Wigner,Zurek,Vitali}. By definition, integrating $W(x,p)$ over a conjugate variable yields the corresponding marginal probability distribution, though it is formally classified as a quasi-probability distribution due to the potential for negative values that highlight its inherently non-classical nature. Figure~\ref{fig6} shows $W(x,p)$ for linear gravitational-like trap strengths $a=0$, $2$, and $10$ under two harmonic confinement strengths: $M=0.5$ [Figs.~\ref{fig6}(a)–\ref{fig6}(c)] and $M=0.25$ [Figs.~\ref{fig6}(d)–\ref{fig6}(f)]. Unless otherwise stated, the parameters are fixed at $t=1$, $\mu=\mu_{0}=E=-2/9$, $G_{1}=-1$, and $G_{2}=0.9999$. For $M=0.5$, increasing $a$ shifts the phase-space distribution away from the origin and broadens its momentum profile, reflecting the momentum imparted by the linear potential. Reducing the harmonic confinement to $M=0.25$ enhances the spatial displacement owing to the weaker restoring force while simultaneously yielding a more localized phase-space distribution. These results demonstrate the interplay between the linear gravitational-like potential and harmonic confinement in controlling the spatial and momentum characteristics of the QDs.

\begin{figure*}
\centering
\includegraphics[scale=.43]{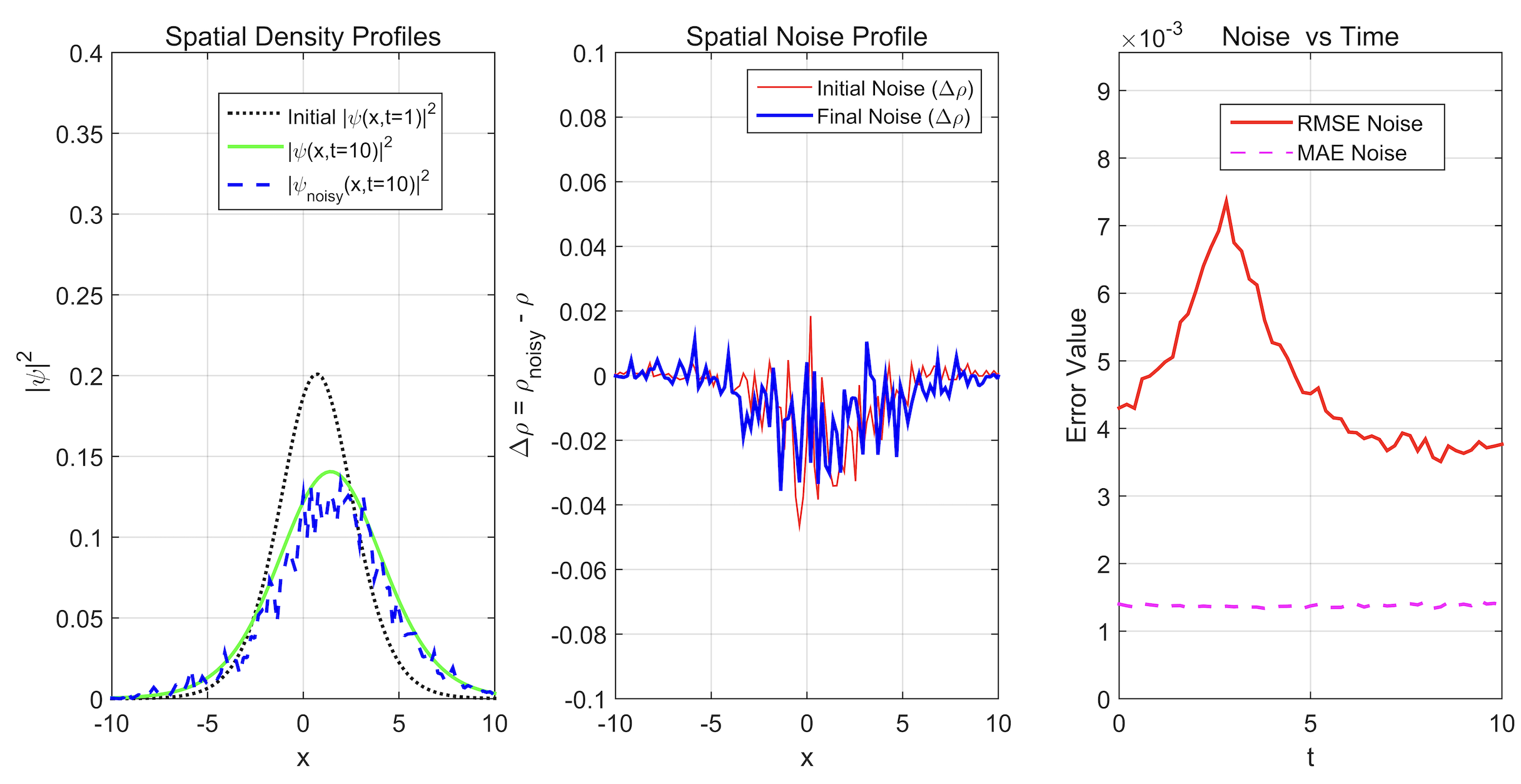}
\caption{\label{stability} Dynamical stability of the analytical quantum droplet solution under weak Gaussian noise ($\eta=0.02$). (a) Density profiles of the unperturbed and perturbed states after real-time propagation using the split-step Fourier method (SSFM). (b) Density difference, $\Delta\rho=\rho_{\mathrm{noisy}}-\rho_c$. (c) Time evolution of the RMSE and MAE, showing bounded errors throughout the propagation and confirming the dynamical stability of the analytical solution. Parameters: $a=0.1$, $M=0.25$, $\mu_0=-2/3$, $E=-2/9$, $G_1=-1$, $G_2=0.999$, $N=1024$, $dx=0.02$, $dt=0.005$, and $t=10$.}
\end{figure*}

\section{Stability Analysis} 
To assess the dynamical stability of the self-bound QD, we perturb the analytical stationary solution of the dimensionless 1D eGPE with complex Gaussian noise,
\begin{equation}
\psi_{\mathrm{noisy}}(x)=\psi(x)+\eta\left[\mathcal{R}{e}(x)+i\mathcal{R}{i}(x)\right],
\end{equation}
where $\eta=0.02$, and $\mathcal{R}{e,i}(x)$ are independent Gaussian random variables with zero mean and unit variance. After adding the perturbation, the wave function is renormalized to conserve the total particle number and propagated in real time using the split-step Fourier method (SSFM). The simulations are performed for $a=0.1$, $M=0.25$, $\mu{0}=-2/3$, $E=-2/9$, $G_{1}=-1$, $G_{2}=0.999$, and $t=1$, using $N=1024$ spatial grid points with $dx=0.02$, a time step $dt=0.005$, and a final evolution time of $t=10$.
The stability is quantified by comparing the noisy and unperturbed density profiles through the root-mean-square error (RMSE),
\begin{equation}
\mathrm{RMSE}(t)=\sqrt{\frac{1}{N}\sum_{i=1}^{N}\left[\rho_{\mathrm{noisy}}(x_i,t)-\rho_c(x_i,t)\right]^2},
\end{equation}
and the mean absolute error (MAE),
\begin{equation}
\mathrm{MAE}(t)=\frac{1}{N}\sum_{i=1}^{N}\left|\rho_{\mathrm{noisy}}(x_i,t)-\rho_c(x_i,t)\right|,
\end{equation}
where $\rho_{\mathrm{noisy}}=|\psi_{\mathrm{noisy}}|^2$ and $\rho_c=|\psi|^2$.
Figure~\ref{stability} shows the density profiles, the corresponding density difference $\Delta\rho=\rho_{\mathrm{noisy}}-\rho_c$, and the temporal evolution of the error metrics. The perturbed and unperturbed density profiles remain nearly indistinguishable throughout the evolution, while the density difference is confined to small-amplitude fluctuations that gradually disperse from the droplet center. The RMSE exhibits a brief initial increase before saturating at a low value ($\sim4\times10^{-3}$), and the MAE remains nearly constant during the propagation. The absence of sustained error growth or density collapse demonstrates that the analytical quantum droplet solution is dynamically stable against weak stochastic perturbations.

\section{Conclusion} In this work, we investigated the dynamics of QDs confined by a harmonic trap in the presence of a constant linear (gravitational-like) potential within the 1D eGPE framework. We obtained exact analytical solutions incorporating both BMF and EMF nonlinearities, recovering the previously known free-space and purely harmonic-trap solutions as limiting cases. The analytical results provide a unified description of the droplet dynamics under combined harmonic and linear confinement. Our analysis reveals two qualitatively distinct dynamical responses. The monopole (breathing) mode remains determined primarily by the harmonic confinement, with its frequency becoming asymptotically insensitive to the linear potential, demonstrating the robustness of the droplet's internal compressional dynamics against uniform external forcing. In contrast, the center-of-mass motion exhibits a pronounced dependence on the competition between the harmonic restoring force and the linear potential. Weak confinement leads to high mobility and large center-of-mass susceptibility, whereas strong confinement suppresses transport even for comparatively large forcing, providing a simple mechanism for tuning droplet transport through the trap frequency. We further showed that the interplay between the harmonic confinement and the linear potential offers a practical means to manipulate the quantum statistical properties of the droplet. A finite linear potential is required to generate enhanced QFI, while stronger harmonic confinement shifts the onset of the high-sensitivity regime to larger forcing strengths and broadens its operational window. The corresponding Wigner phase-space distributions reveal that weaker confinement enhances spatial displacement while producing a more localized phase-space structure, highlighting the close connection between external confinement, transport, and quantum-state engineering. Numerical simulations based on the split-step Fourier method confirm the dynamical stability of the analytical solutions. 

In summary, the present investigations show that a constant linear gravitational like trap provides a versatile platform for controlling the transport and quantum metrological properties of ultradilute 1D QDs while leaving their intrinsic breathing dynamics largely unaffected. The analytical framework developed here can be extended to higher-dimensional geometries, time-dependent or spatially varying external fields, and spinor or dipolar quantum droplets, where additional non-equilibrium phenomena and collective excitations may arise. These extensions provide a natural direction for future theoretical and experimental studies of controlled transport and quantum sensing with self-bound Bose mixtures.

\end{document}